\documentclass[prl,twocolumn,notitlepage,superscriptaddress,longbibliography]{revtex4-1}
\usepackage{amsmath}
\usepackage{amssymb}

\usepackage[unicode=true,colorlinks=true,citecolor=blue,urlcolor=blue]{hyperref}

\usepackage{bm}
\usepackage{epsfig}
\usepackage{multirow}

\usepackage[normalem]{ulem}
\renewcommand {\Im}{\mathop\mathrm{Im}\nolimits}
\renewcommand {\Re}{\mathop\mathrm{Re}\nolimits}

\renewcommand {\phi}{{\varphi}}
\newcommand {\rmi}{{\rm i}}

\newcommand {\e}{{\rm e}}
\newcommand {\eps}{\varepsilon}

\newcommand{\gf}{\gamma_{\to}}
\newcommand{\gb}{\gamma_{\leftarrow}}
\begin{document}
\title{Mesoscopic non-Hermitian skin effect}

\author{Alexander Poddubny}
\email{poddubny@weizmann.ac.il}

\affiliation{Department of Physics of Complex Systems, Weizmann Institute of Science, Rehovot 7610001, Israel}

\author{Janet Zhong}
\email{janetzh@stanford.edu}
\author{Shanhui Fan}
\affiliation{Department of Applied Physics, Stanford University, Stanford, California 94305, USA}

%

\begin{abstract}
We discuss a  generalization of the non-Hermitian skin effect to finite-size photonic structures with
 neither gain nor loss in the bulk and purely real energy spectrum under periodic boundary conditions (PBC).
 We show that  such systems can still have significant portions of eigenmodes concentrated at the edges and  that this edge concentration can be linked to the non-trivial point-gap topology of the size-dependent regularized PBC spectrum, accounting for the  radiative losses. As an example, we consider the chiral waveguide quantum electrodynamics platform with an array of atoms coupled to the waveguide. The proposed mesoscopic analogue of the non-Hermitian skin effect could be potentially applied to other  seemingly lossless photonic structures, such as chiral resonant all-dielectric metamaterials.
\end{abstract}
\maketitle

\begin{figure}[b]
\centering\includegraphics[width=0.48\textwidth]{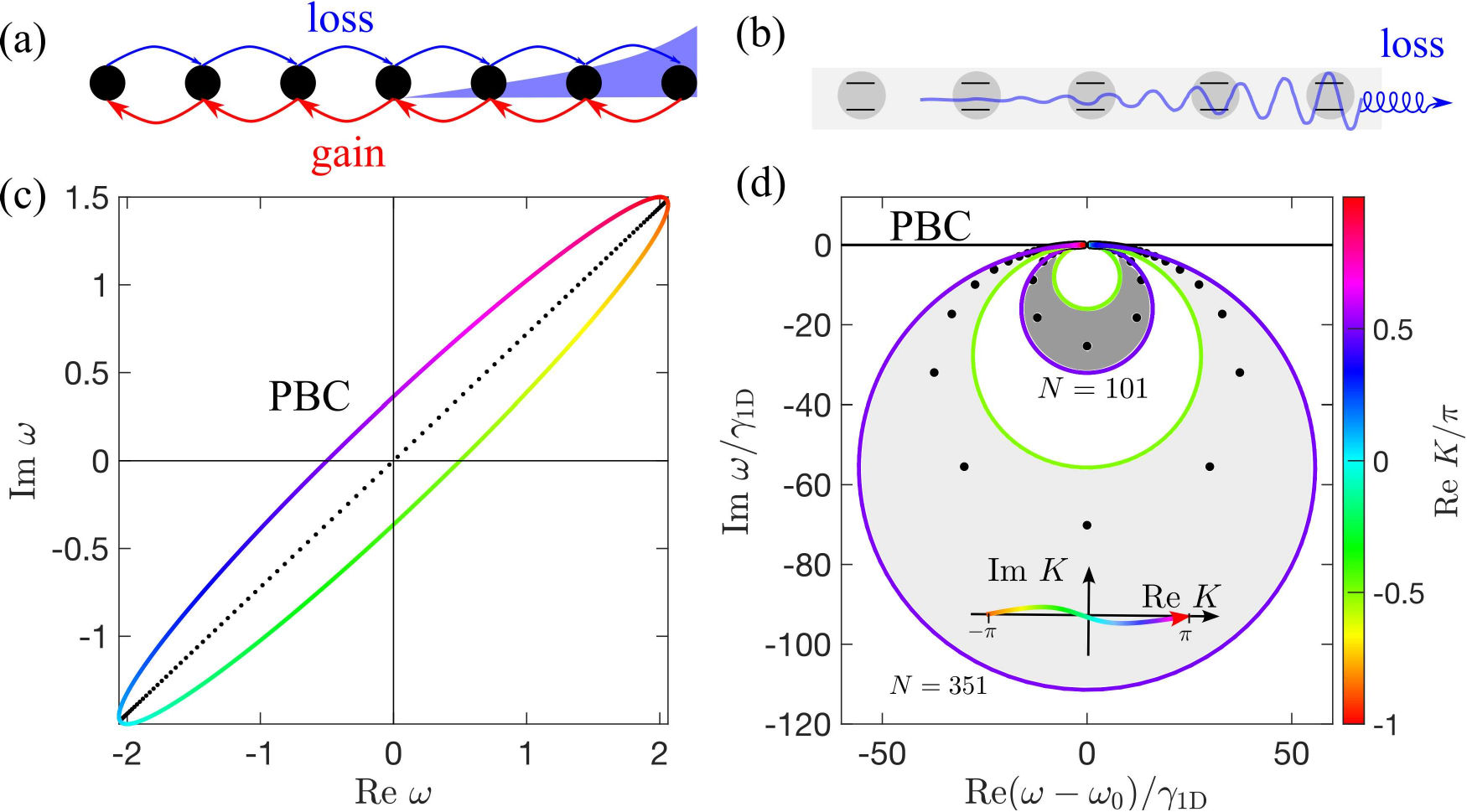}
\caption{(a,b) Schematics of (a) Hatano-Nelson model with spatially distributed gain and loss and (b) chiral waveguide-QED model with radiative loss on the edges. (c,d) spectra under periodic boundary conditions (PBC, colored lines) and open boundary conditions (OBC,  black dots) in these models. 
Hatano-Nelson model corresponded to $t_1=1+0.5\rmi$, $t_2=1+\rmi$ and $N=102$ sites.  Two horseshoe-shaped lines in (d) show regularized PBCs for $N=101$ and $N=351$ atoms. Inset illustrates the complex $K$ contour used for regularized PBC. Calculation has been performed for $\xi=1/2$ and $\varphi=\pi/2$.
}\label{fig:1}
\end{figure}

{\it Introduction.} 
Non-Hermitian skin effect (NHSE) is manifested by the concentration of the eigenmodes at the edge of a finite structure under open boundary conditions (OBC), that is related to the nontrivial point-gap topology under periodic BC~\cite{Lee2016,Torres2018,Kunst2018,Yao2018,Slager2020,Okuma2020,Bergholtz2021}. 
Particular realizations of  NHSE can be very different, such as tight-binding lattices, photonic crystals~\cite{Zhong2021NH,Longhi2021},  continuous media~\cite{Yokomizo2022}, and even lattices in synthetic dimensions~\cite{Wang_2021}. In all these cases, the energy spectrum drastically changes under the open boundary conditions (OBC), contrary to the usual Hermitian systems. A paradigmatic example is the Hatano-Nelson model  of one-dimensional lattice with different left- and right-tunneling coupling constants $t_1\ne t_2$~\cite{Hatano1997}, see Fig.~\ref{fig:1}(a). Its PBC spectrum has a loop in the complex plane with a non-zero winding number, that under OBC collapses into a line corresponding to edge-concentrated eigenmodes. 

Here we apply this NHSE perspective to the edge eigenmode concentration in  a  very different photonic system, that has purely real PBC spectrum and neither loss nor gain in the bulk, see Fig.~\ref{fig:1}(b,d). When being put under open boundary conditions, a finite structure still acquires radiative losses due to the photon escape into the far field, so the OBC  spectrum becomes complex. These OBC eigenmodes can also concentrate at one edge when the structure is chiral. We aim to link  this concentration to the non-trivial point-gap topology of dispersion law $\omega(K)$ for complex wave vectors  $K$. To this end, we propose a   regularization of the  $\omega(K)$ dependence for a finite structure. The real-valued PBC spectrum is dramatically modified by this regularization. It becomes complex and encircles the OBC spectrum, see horseshoe contours in Fig.~\ref{fig:1}(d), just like in the usual NHSE case of Fig.~\ref{fig:1}(c). In stark contrast to the usual scenario, the regularized PBC depends on the system size, so the considered NHSE has a mesoscopic nature.

We consider a particular example of mesoscopic NHSE for the chiral waveguide quantum electrodynamics (QED) setup, that is, for an array of natural or artificial atoms, chirally coupled to the waveguide~\cite{Lodahl2017,Sheremet2023}. Such arrays are now studied both experimentally \cite{Prasad2020,Liedl_2023,du2023giant} and theoretically. While they are known to have eigenmodes concentrated at the edges~\cite{Kornovan2017,Fedorovich2022,Yang_2022,Wang_2022,Zeng2022,Zeng2023}, we believe that  the size-dependent connection to the NHSE  is not yet fully understood. In particular, while Ref.~\cite{Fedorovich2022}, with one of us as a coauthor, analyzed the eigenmodes in great detail, this work has not mentioned any topological spectral features.  Refs.~\cite{Zeng2022,Zeng2023} considered a very similar setup for magnons that, however,  had an inherent nonzero nonradiative loss.
Refs.~\cite{Yang_2022,Wang_2022} did discuss NHSE in substantially different chiral atomic systems, where the main loss mechanism was due to the emission into the free space~\cite{Yang_2022}, i.e. perpendicular to the waveguide or perpendicular to the plane of the atomic mirror ~\cite{Wang_2022}, rather than at the edges. Thus, already the non-regularized PBC spectrum in Refs.~\cite{Zeng2022,Zeng2023,Yang_2022,Wang_2022} has been complex corresponding to the usual NHSE case of Fig.~\ref{fig:1}(a,c) rather than to the proposed mescoscopic NHSE.

{\it Model}. We consider single-excited eigenstates in an array of $N$ periodically spaced emitters described by the effective non-Hermitian Hamiltonian~\cite{Kornovan2017}
$H=\sum_{m,n=1}^NH_{m,n}\sigma_n^\dag\sigma_m^{\vphantom{\dag}} $ with 
$\sigma_m^\dag$ being the raising operators and 
\begin{equation}\label{eq:H}
H_{m,n}=\omega_0\delta_{m,n}-\rmi\begin{cases}
 \gf \e^{\rmi\varphi |m-n|}, &m>n\\
 \frac{\gf+\gb}{2}, &m=n\\
 \gb \e^{\rmi\varphi |m-n|}, &m<n
\end{cases}\:,
\end{equation}
where 
$ \gf=2\gamma_{\rm 1D}/(1+\xi),$  $ \gb=2\gamma_{\rm 1D}\xi/(1+\xi)$, 
are the spontaneous emission rates into the waveguide in the forward and backward directions. 
The phase $\varphi=\omega_0d/c$
 is the phase gained by light between the two emitters. We use the Markovian approximation,  valid for $\gamma_{\rm 1D}\ll \omega$, so that the frequency dependence of $\varphi$ is ignored.
Notably, the Hamiltonian Eq.~\eqref{eq:H} features long-range photon-mediated coupling between distant emitters. 
The parameter $\xi$ is the ratio of emission rates in the forward and backward directions.
Under the PBC, the  Hamiltonian matrix $H_{mn}$ has the eigenstates $\psi_m=\e^{\rmi Km}$
with the eigenenergies
\begin{equation}\label{eq:disp1}
\omega(K)=\omega_0+\gamma_{\rm 1D}\frac{\sin\varphi+c\sin K}{\cos K-\cos\varphi},\text{ where } 
c=\frac{1-\xi}{1+\xi}\:.
\end{equation}
Importantly, the energy spectrum Eq.~\eqref{eq:disp1} is purely real. It features a singularity near the light line, $K=\varphi$, and a polaritonic band gap near the emitter resonance frequency  $\omega_0$. We can also rewrite the dispersion law  in the effective medium approximation, assuming $\varphi\ll 1$, $K\ll 1$, that leads to $K^2=\varphi^2\eps(\omega,K)$ where 
\begin{equation}\label{eq:eps}
\eps(\omega,K)=1+\frac{2\gamma_{\rm 1D}}{\varphi}\frac{1+cK}{\omega_0-\omega}
\end{equation}
 is the resonant permittivity. The breakdown of the time-reversal symmetry for $c\ne 0$ is manifested by the linear-in-$K$ term in Eq.~\eqref{eq:eps}. 

The OBC spectrum  can be obtained either directly by diagonalizing Eq.~\eqref{eq:H} or by solving the equation \cite{Fedorovich2022,Voronov2007JLu}
\begin{equation}\label{eq:FP} 
r_{\hookrightarrow}(\omega) r_{\hookleftarrow}(\omega)\e^{\rmi (K_+-K_-)(N-1)}=1
\end{equation}
where 
$
r_{\hookrightarrow/\hookleftarrow}=-(\e^{\pm\rmi K_{\pm}}-\e^{\rmi \varphi})/(\e^{\pm\rmi K_{\mp}}-\e^{\rmi \varphi})$
are the reflection coefficients of the polaritonic modes from the inside of the structure and 
$K_{\pm}$ are the solutions for the dispersion equation for forward and backward propagating waves.
 Equation~\eqref{eq:FP} describes standing waves in the cavity made of resonant chiral material with the (effective) permittivity Eq.~\eqref{eq:eps}. The chiral WQED is related to the non-reciprocal waveguide system
\cite{Tsakmakidis_2017}, that has attracted a lot of interest \cite{Mann2019,Tsakmakidis2020,Mann_2020,Buddhiraju_2020}.

\begin{figure}[t]
\centering\includegraphics[width=0.48\textwidth]{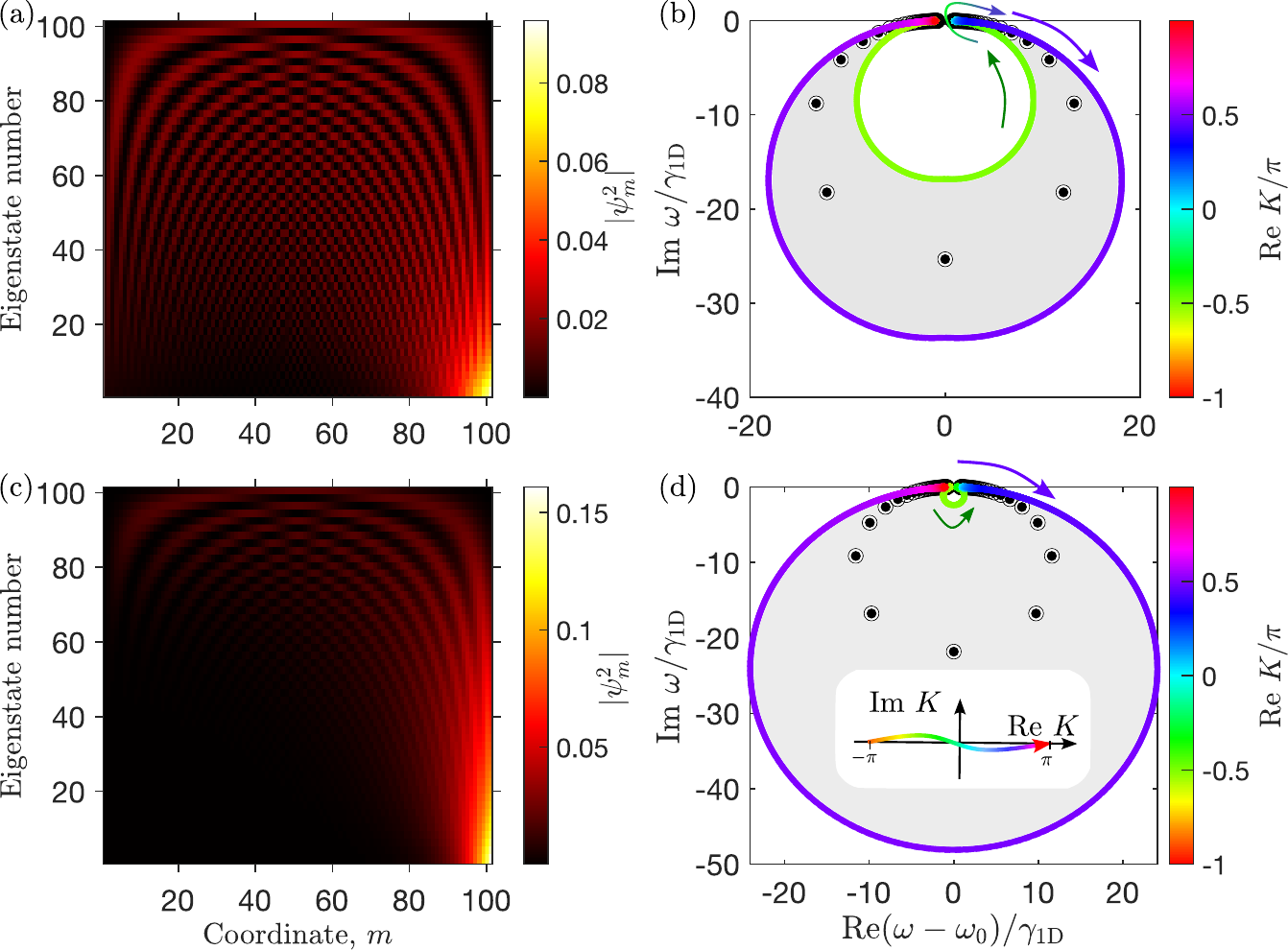}
\caption{ 
Eigenmodes (a,c) and energy spectrum (c,d) of the structure with $N=101$ emitters and $\xi=0.5$ (a,b) and 
and $\xi=0.05$ (c,d).  Filled circles correspond to the numerical solution, open circles are calculated according to Eq.~\eqref{eq:Lambert}.
The eigenmodes in (a,b) are presented in the descending order of the radiative decay rate. 
The PBC spectrum is shown by colored lines in (c,d).
Inset in (d) schematically illustrates the complex $K$ contour used for the PBC calculation.
}
\label{fig:2}
\end{figure}

{\it Mescoscopic NHSE.}
Figure~\ref{fig:2} presents the eigenstates and the OBC spectra calculated for the finite arrays with $N=101$ emitters for two different asymmetry parameters $\xi=0.5$ (a,b) and 
and $\xi=0.05$ (c,d). In both cases, the structures feature superradiant modes with a much larger decay rate than the single atom decay rate $\gamma_{\rm 1D}$~\cite{Fedorovich2022}. 
 
The eigenmodes in Fig.~\ref{fig:2}(a,c) are ordered by the decreasing radiative decay rate and it is clear that they become pinned to  the edges for higher decay rate. In the more chiral case, Fig.~\ref{fig:2}(c), the large fraction of eigenmodes is concentrated at just one edge, which looks like  the hallmark of the NHSE.  
Indeed,  increasing the imaginary part of $\omega$ suppresses the resonant reflection coefficients $r_{\hookrightarrow/\hookleftarrow}$ which  means easier photon escape outside of the structure. At the same time, the difference between $\Im K_+$ and $\Im K_-$ increases in the presence of the chirality. Given  the  eigenmode spatial profile~\cite{Fedorovich2022}
\begin{multline}\label{eq:OBC}
\psi_m\propto\e^{\rmi K_+(m-N)}+r_{\hookleftarrow}\e^{\rmi K_-(m-N)} 
\\\propto\e^{\rmi K_+(m-1)}r_{\hookrightarrow}+\e^{\rmi K_-(m-1)}\:,
\end{multline}
the increase of this difference means localization of the eigenmodes at one particular edge. 
For $|\omega-\omega_0|\gg \gamma_{\rm 1D}$ and $\varphi=\pi/2$ we can obtain  from Eq.~\eqref{eq:FP} an approximate analytical expression for the OBC spectrum
 \begin{equation}\label{eq:Lambert}
\omega_\nu^{\pm}-\omega_0=-\frac{\rmi N\gamma_{\rm 1D}}{W_{\nu}\left(\pm2N/\sqrt{1-c^2}\right)},\quad \nu=0,\pm 1,\pm 2\ldots
\end{equation}
of the most superradiant modes, which is shown by open circles in Fig.~\ref{fig:2}(b,d) and well describes the numerical results. Here, $W_\nu(z)$ is the Lambert function, satisfying the equation $W_{\nu}\e^{W_{\nu}}=z$.
 For $N\gg 1$ we can  approximate it as $W(z)\approx \ln z-\ln\ln z-\rmi \pi\nu z$, which makes evident a weak logarithmic suppression of the linear growth of the radiative decay rate with $N$. Such suppression is distinct from the usual single superradiant Dicke mode for $\varphi=0$, see also Ref.~\cite{Wang_2023} for more detail on the $\varphi=0$ case.
\begin{figure}[t]
\centering\includegraphics[width=0.49\textwidth]{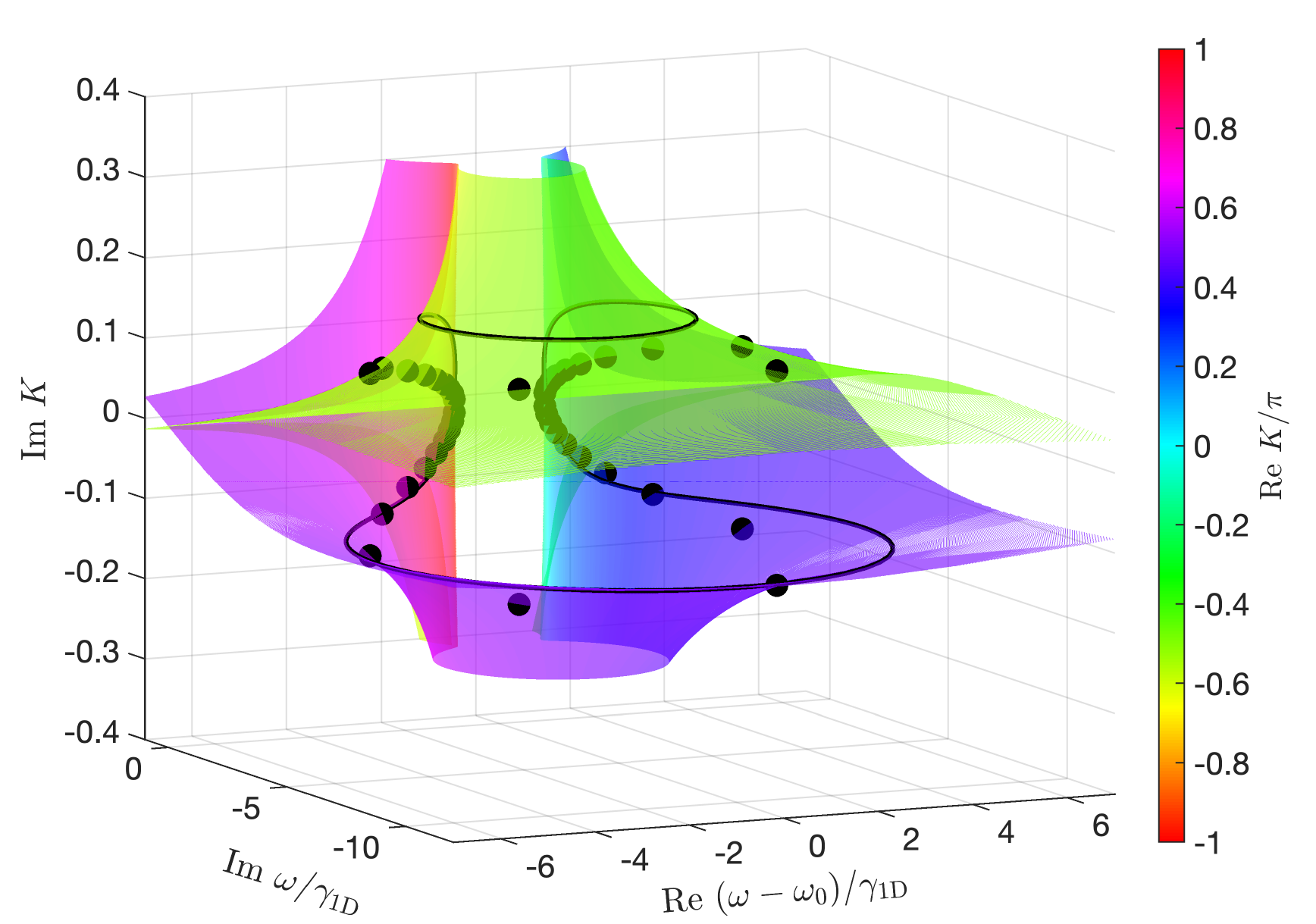}
\caption{ Riemann surface of the dispersion law $\omega(K)$. Color indicates the value of $\Re K$. Dots show the values of $K_\pm$ for the eigenfrequencies of the finite structure $\omega$. Thick black line corresponds to the contour $K=\Re K-\rmi 0.13\pi\sin \Re K.$   Calculation has been performed for $\xi=0.5$ ($c=1/3$), $\varphi=\pi/2$  and $N=20$.
}\label{fig:Riemann}
\end{figure}


The presence of edge-concentrated eigenmodes in the finite structures has inspired us for more detailed analysis of the PBC energy spectra. {The $N\to \infty$ limit of infinite structure with purely zero loss is not well defined.} Even in vacuum,  the photon propagator $1/[(\omega/c)^2-K^2]$  has a  singularity in $K$ space that has to be carefully treated if one needs for example, to use the Cauchy theorem to  integrate over $K$ or $\omega$~\cite{berestetskii1982quantum}. Here we consider the excitations
with $\Im\omega<0$ that decay in time  $\propto \exp(-\rmi\omega t)$, and one should add the imaginary part to the wave vector $K$ with an appropriate sign. We propose the regularization
\begin{equation}\label{eq:reg}
K\to \Re K-\rmi \delta K(N) \sin \Re K,-\pi\le\Re K<\pi
\end{equation}
for the dispersion law Eq.~\eqref{eq:disp1}. The small parameter $\delta K(N)$ accounts for the finite (radiative) losses  that render the PBC spectrum Eq.~\eqref{eq:disp1} complex.
The horseshoe-like contour of the regularized PBC is shown in Fig.~\ref{fig:1}(a) and Fig.~\ref{fig:2}(b,d) by  lines with gradient color that encodes the value of $\Re K$. This regularized PBC spectrum has a loop around the OBC  characterized by the nonvanishing winding number. 
 For vanishing chirality, when $\omega(K)=\omega(-K)$, the horseshoe shrinks to a line with zero area and for larger chirality it expands to a full circle (Fig.~\ref{fig:2}d).  
 Contrary to the usual NHSE case, the regularization parameter $\delta K$ in  Eq.~\eqref{eq:reg} is not universal. It depends on $N$ and tends to zero for $N\to \infty$.  The curves in Fig.~\ref{fig:1}(d) and  Fig.~\ref{fig:2} have been calculated for {$\delta K=4/N$}. This scaling describes how for larger $N$ the horseshoe expands and at the same time the ends of the horseshoe approach the real-valued PBC spectrum, see Fig.~\ref{fig:1}(d). Numerical analysis hints at a  more complicated scaling  $\delta K\propto \ln N/N$, which reflects the dependence of the lifetime of the modes Eq.~\eqref{eq:Lambert} on $N$. 
 
 It is also instructive to analyze  the Riemann surface of $\omega(K)$ that is shown in Fig.~\ref{fig:Riemann}  depending on both  $\Re K$ and $\Im K$ together with the OBC energy spectrum and the regularization path Eq.~\eqref{eq:reg}. The color of the Riemann surface encodes the value of $\Re K$. In such a way, it becomes evident that the non-zero winding is the generic feature of the complex energy spectrum of the chiral structure that is independent of the particular regularization. On the other hand, Fig.~\ref{fig:Riemann}  also demonstrates a clear distinction from the usual systems with NHSE: the signs of $\Im K_{+}$ and $\Im K_{-}$ for the OBC eigenmodes are opposite. The eigenmode localization at one particular edge, apparent in Fig.~\ref{fig:2}(c), is just due to the different weight of the terms $\propto\e^{\rmi K_\pm m}$ in Eq.~\eqref{eq:OBC} and not because both exponents decay into the same direction.

\begin{figure}[t]
\centering\includegraphics[width=0.48\textwidth]{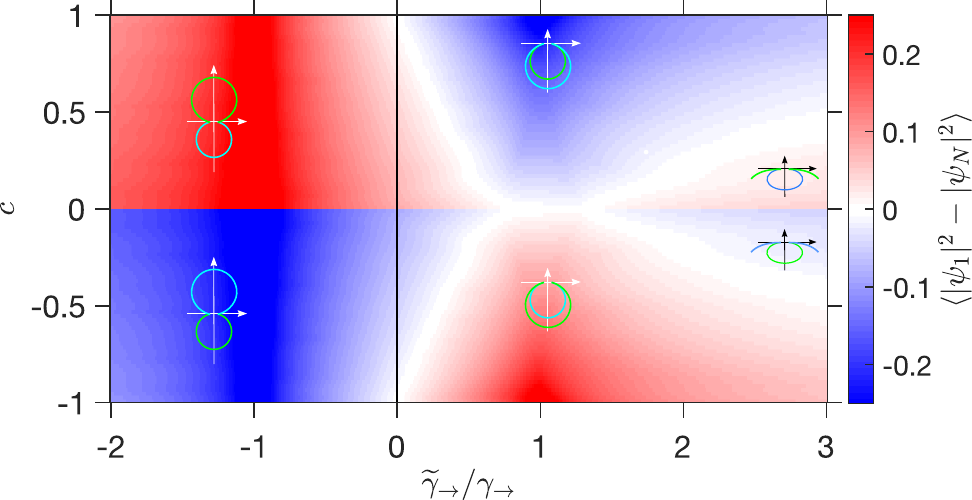}
\caption{ Localization of the eigenmodes at the edges depending on the   loss parameter $\widetilde \gamma_{\to}$ and the chirality degree $c$. The insets schematically illustrate the regularized PBC spectra.
The calculation has been performed for $\varphi=\pi/2$  and $N=20$.
}\label{fig:phases}
\end{figure}
The control of the mesoscopic NHSE  by the  losses or gain through the edges can be explicitly demonstrated by analyzing the non-Hermitian part of the Hamiltonian Eq.~\eqref{eq:H} 
\[
H_{nm}-H_{mn}^{*}=-\rmi\gamma_{\to}\e^{\rmi \varphi(n-m)}-\rmi\gamma_{\leftarrow}\e^{\rmi \varphi(m-n)}\:.
\] The presence of just two terms in the singular value decomposition of $H-H^{\dag}$ is yet another manifestation of the unusual nature of considered non-Hermiticity;  they describe  emission to the right and to the left~\cite{Cardenas_Lopez_2023}.
It is  instructive to explicitly modify the $\gamma_{\to}$ coefficient in the non-Hermitian part,
$\gamma_{\to}\to\widetilde \gamma_{\to}$.
We assume this can be realized by controlling the medium outside the structure. The results are presented in Fig.~\ref{fig:phases}, where the color encodes the edge localization degree $|\psi_1|^2-|\psi_N|^2$, averaged over all the eigenstates $\psi$. For completeness, we also considered the case $\widetilde\gamma_{\to}<0$, which means the presence of the gain.  Importantly, the usual PBC spectrum does not depend on $\widetilde \gamma_{\to}$, while the regularized PBC does (see the insets in Fig.~\ref{fig:phases}). Thus,  Fig.~\ref{fig:phases} demonstrates that despite the usual PBC spectrum staying real, the modification of the boundary conditions can control the sign of the mesoscopic NHSE by switching the mode localization direction in the finite array.

{\it Summary}. To summarize, we show how the concept of nontrivial point-gap topology, usually used under the periodic boundary conditions, can be used to analyze a mesoscopic analog of the non-Hermitian skin effect 
in a finite-size photonic structure with the only loss or gain being at the edges. While our consideration has been limited to the simplest one-dimensional model, it also hints at possible eigenmode concentration at the edges of various resonant structures, provided they are chiral and lossless inside. These could be, for example,  chiral all-dielectric metasurfaces with Mie resonances~\cite{Kivshar2022,Koshelev2023}. 

{\it Acknowledgments}. 
The work of ANP has been supported by  research grants from the Center for New Scientists 
and from the Center for Scientific Excellence at the Weizmann Institute of Science. S. F. acknowledges a Simons Investigator in Physics
grant from the Simons Foundation (Grant No. 827065)

\begin{thebibliography}{34}%
\makeatletter
\providecommand \@ifxundefined [1]{%
 \@ifx{#1\undefined}
}%
\providecommand \@ifnum [1]{%
 \ifnum #1\expandafter \@firstoftwo
 \else \expandafter \@secondoftwo
 \fi
}%
\providecommand \@ifx [1]{%
 \ifx #1\expandafter \@firstoftwo
 \else \expandafter \@secondoftwo
 \fi
}%
\providecommand \natexlab [1]{#1}%
\providecommand \enquote  [1]{``#1''}%
\providecommand \bibnamefont  [1]{#1}%
\providecommand \bibfnamefont [1]{#1}%
\providecommand \citenamefont [1]{#1}%
\providecommand \href@noop [0]{\@secondoftwo}%
\providecommand \href [0]{\begingroup \@sanitize@url \@href}%
\providecommand \@href[1]{\@@startlink{#1}\@@href}%
\providecommand \@@href[1]{\endgroup#1\@@endlink}%
\providecommand \@sanitize@url [0]{\catcode `\\12\catcode `\$12\catcode
  `\&12\catcode `\#12\catcode `\^12\catcode `\_12\catcode `\%12\relax}%
\providecommand \@@startlink[1]{}%
\providecommand \@@endlink[0]{}%
\providecommand \url  [0]{\begingroup\@sanitize@url \@url }%
\providecommand \@url [1]{\endgroup\@href {#1}{\urlprefix }}%
\providecommand \urlprefix  [0]{URL }%
\providecommand \Eprint [0]{\href }%
\providecommand \doibase [0]{http://dx.doi.org/}%
\providecommand \selectlanguage [0]{\@gobble}%
\providecommand \bibinfo  [0]{\@secondoftwo}%
\providecommand \bibfield  [0]{\@secondoftwo}%
\providecommand \translation [1]{[#1]}%
\providecommand \BibitemOpen [0]{}%
\providecommand \bibitemStop [0]{}%
\providecommand \bibitemNoStop [0]{.\EOS\space}%
\providecommand \EOS [0]{\spacefactor3000\relax}%
\providecommand \BibitemShut  [1]{\csname bibitem#1\endcsname}%
\let\auto@bib@innerbib\@empty
\bibitem [{\citenamefont {Lee}(2016)}]{Lee2016}%
  \BibitemOpen
  \bibfield  {author} {\bibinfo {author} {\bibfnamefont {Tony~E.}\ \bibnamefont
  {Lee}},\ }\bibfield  {title} {\enquote {\bibinfo {title} {Anomalous edge
  state in a non-{H}ermitian lattice},}\ }\href {\doibase
  10.1103/PhysRevLett.116.133903} {\bibfield  {journal} {\bibinfo  {journal}
  {Phys. Rev. Lett.}\ }\textbf {\bibinfo {volume} {116}},\ \bibinfo {pages}
  {133903} (\bibinfo {year} {2016})}\BibitemShut {NoStop}%
\bibitem [{\citenamefont {Martinez~Alvarez}\ \emph {et~al.}(2018)\citenamefont
  {Martinez~Alvarez}, \citenamefont {Barrios~Vargas},\ and\ \citenamefont
  {Foa~Torres}}]{Torres2018}%
  \BibitemOpen
  \bibfield  {author} {\bibinfo {author} {\bibfnamefont {V.~M.}\ \bibnamefont
  {Martinez~Alvarez}}, \bibinfo {author} {\bibfnamefont {J.~E.}\ \bibnamefont
  {Barrios~Vargas}}, \ and\ \bibinfo {author} {\bibfnamefont {L.~E.~F.}\
  \bibnamefont {Foa~Torres}},\ }\bibfield  {title} {\enquote {\bibinfo {title}
  {Non-{H}ermitian robust edge states in one dimension: Anomalous localization
  and eigenspace condensation at exceptional points},}\ }\href {\doibase
  10.1103/PhysRevB.97.121401} {\bibfield  {journal} {\bibinfo  {journal} {Phys.
  Rev. B}\ }\textbf {\bibinfo {volume} {97}},\ \bibinfo {pages} {121401}
  (\bibinfo {year} {2018})}\BibitemShut {NoStop}%
\bibitem [{\citenamefont {Kunst}\ \emph {et~al.}(2018)\citenamefont {Kunst},
  \citenamefont {Edvardsson}, \citenamefont {Budich},\ and\ \citenamefont
  {Bergholtz}}]{Kunst2018}%
  \BibitemOpen
  \bibfield  {author} {\bibinfo {author} {\bibfnamefont {Flore~K.}\
  \bibnamefont {Kunst}}, \bibinfo {author} {\bibfnamefont {Elisabet}\
  \bibnamefont {Edvardsson}}, \bibinfo {author} {\bibfnamefont {Jan~Carl}\
  \bibnamefont {Budich}}, \ and\ \bibinfo {author} {\bibfnamefont {Emil~J.}\
  \bibnamefont {Bergholtz}},\ }\bibfield  {title} {\enquote {\bibinfo {title}
  {Biorthogonal bulk-boundary correspondence in non-{H}ermitian systems},}\
  }\href {\doibase 10.1103/PhysRevLett.121.026808} {\bibfield  {journal}
  {\bibinfo  {journal} {Phys. Rev. Lett.}\ }\textbf {\bibinfo {volume} {121}},\
  \bibinfo {pages} {026808} (\bibinfo {year} {2018})}\BibitemShut {NoStop}%
\bibitem [{\citenamefont {Yao}\ and\ \citenamefont {Wang}(2018)}]{Yao2018}%
  \BibitemOpen
  \bibfield  {author} {\bibinfo {author} {\bibfnamefont {Shunyu}\ \bibnamefont
  {Yao}}\ and\ \bibinfo {author} {\bibfnamefont {Zhong}\ \bibnamefont {Wang}},\
  }\bibfield  {title} {\enquote {\bibinfo {title} {Edge states and topological
  invariants of non-{H}ermitian systems},}\ }\href {\doibase
  10.1103/PhysRevLett.121.086803} {\bibfield  {journal} {\bibinfo  {journal}
  {Phys. Rev. Lett.}\ }\textbf {\bibinfo {volume} {121}},\ \bibinfo {pages}
  {086803} (\bibinfo {year} {2018})}\BibitemShut {NoStop}%
\bibitem [{\citenamefont {Borgnia}\ \emph {et~al.}(2020)\citenamefont
  {Borgnia}, \citenamefont {Kruchkov},\ and\ \citenamefont
  {Slager}}]{Slager2020}%
  \BibitemOpen
  \bibfield  {author} {\bibinfo {author} {\bibfnamefont {Dan~S.}\ \bibnamefont
  {Borgnia}}, \bibinfo {author} {\bibfnamefont {Alex~Jura}\ \bibnamefont
  {Kruchkov}}, \ and\ \bibinfo {author} {\bibfnamefont {Robert-Jan}\
  \bibnamefont {Slager}},\ }\bibfield  {title} {\enquote {\bibinfo {title}
  {Non-{H}ermitian boundary modes and topology},}\ }\href {\doibase
  10.1103/PhysRevLett.124.056802} {\bibfield  {journal} {\bibinfo  {journal}
  {Phys. Rev. Lett.}\ }\textbf {\bibinfo {volume} {124}},\ \bibinfo {pages}
  {056802} (\bibinfo {year} {2020})}\BibitemShut {NoStop}%
\bibitem [{\citenamefont {Okuma}\ \emph {et~al.}(2020)\citenamefont {Okuma},
  \citenamefont {Kawabata}, \citenamefont {Shiozaki},\ and\ \citenamefont
  {Sato}}]{Okuma2020}%
  \BibitemOpen
  \bibfield  {author} {\bibinfo {author} {\bibfnamefont {Nobuyuki}\
  \bibnamefont {Okuma}}, \bibinfo {author} {\bibfnamefont {Kohei}\ \bibnamefont
  {Kawabata}}, \bibinfo {author} {\bibfnamefont {Ken}\ \bibnamefont
  {Shiozaki}}, \ and\ \bibinfo {author} {\bibfnamefont {Masatoshi}\
  \bibnamefont {Sato}},\ }\bibfield  {title} {\enquote {\bibinfo {title}
  {Topological origin of non-{H}ermitian skin effects},}\ }\href {\doibase
  10.1103/PhysRevLett.124.086801} {\bibfield  {journal} {\bibinfo  {journal}
  {Phys. Rev. Lett.}\ }\textbf {\bibinfo {volume} {124}},\ \bibinfo {pages}
  {086801} (\bibinfo {year} {2020})}\BibitemShut {NoStop}%
\bibitem [{\citenamefont {Bergholtz}\ \emph {et~al.}(2021)\citenamefont
  {Bergholtz}, \citenamefont {Budich},\ and\ \citenamefont
  {Kunst}}]{Bergholtz2021}%
  \BibitemOpen
  \bibfield  {author} {\bibinfo {author} {\bibfnamefont {Emil~J.}\ \bibnamefont
  {Bergholtz}}, \bibinfo {author} {\bibfnamefont {Jan~Carl}\ \bibnamefont
  {Budich}}, \ and\ \bibinfo {author} {\bibfnamefont {Flore~K.}\ \bibnamefont
  {Kunst}},\ }\bibfield  {title} {\enquote {\bibinfo {title} {Exceptional
  topology of non-{H}ermitian systems},}\ }\href {\doibase
  10.1103/RevModPhys.93.015005} {\bibfield  {journal} {\bibinfo  {journal}
  {Rev. Mod. Phys.}\ }\textbf {\bibinfo {volume} {93}},\ \bibinfo {pages}
  {015005} (\bibinfo {year} {2021})}\BibitemShut {NoStop}%
\bibitem [{\citenamefont {Zhong}\ \emph {et~al.}(2021)\citenamefont {Zhong},
  \citenamefont {Wang}, \citenamefont {Park}, \citenamefont {Asadchy},
  \citenamefont {Wojcik}, \citenamefont {Dutt},\ and\ \citenamefont
  {Fan}}]{Zhong2021NH}%
  \BibitemOpen
  \bibfield  {author} {\bibinfo {author} {\bibfnamefont {Janet}\ \bibnamefont
  {Zhong}}, \bibinfo {author} {\bibfnamefont {Kai}\ \bibnamefont {Wang}},
  \bibinfo {author} {\bibfnamefont {Yubin}\ \bibnamefont {Park}}, \bibinfo
  {author} {\bibfnamefont {Viktar}\ \bibnamefont {Asadchy}}, \bibinfo {author}
  {\bibfnamefont {Charles~C.}\ \bibnamefont {Wojcik}}, \bibinfo {author}
  {\bibfnamefont {Avik}\ \bibnamefont {Dutt}}, \ and\ \bibinfo {author}
  {\bibfnamefont {Shanhui}\ \bibnamefont {Fan}},\ }\bibfield  {title} {\enquote
  {\bibinfo {title} {Nontrivial point-gap topology and non-{H}ermitian skin
  effect in photonic crystals},}\ }\href {\doibase 10.1103/PhysRevB.104.125416}
  {\bibfield  {journal} {\bibinfo  {journal} {Phys. Rev. B}\ }\textbf {\bibinfo
  {volume} {104}},\ \bibinfo {pages} {125416} (\bibinfo {year}
  {2021})}\BibitemShut {NoStop}%
\bibitem [{\citenamefont {Longhi}(2021)}]{Longhi2021}%
  \BibitemOpen
  \bibfield  {author} {\bibinfo {author} {\bibfnamefont {Stefano}\ \bibnamefont
  {Longhi}},\ }\bibfield  {title} {\enquote {\bibinfo {title} {Non-{H}ermitian
  skin effect beyond the tight-binding models},}\ }\href {\doibase
  10.1103/physrevb.104.125109} {\bibfield  {journal} {\bibinfo  {journal}
  {Phys. Rev. B}\ }\textbf {\bibinfo {volume} {104}},\ \bibinfo {pages}
  {125109} (\bibinfo {year} {2021})}\BibitemShut {NoStop}%
\bibitem [{\citenamefont {Yokomizo}\ \emph {et~al.}(2022)\citenamefont
  {Yokomizo}, \citenamefont {Yoda},\ and\ \citenamefont
  {Murakami}}]{Yokomizo2022}%
  \BibitemOpen
  \bibfield  {author} {\bibinfo {author} {\bibfnamefont {Kazuki}\ \bibnamefont
  {Yokomizo}}, \bibinfo {author} {\bibfnamefont {Taiki}\ \bibnamefont {Yoda}},
  \ and\ \bibinfo {author} {\bibfnamefont {Shuichi}\ \bibnamefont {Murakami}},\
  }\bibfield  {title} {\enquote {\bibinfo {title} {Non-{H}ermitian waves in a
  continuous periodic model and application to photonic crystals},}\ }\href
  {\doibase 10.1103/physrevresearch.4.023089} {\bibfield  {journal} {\bibinfo
  {journal} {Phys. Rev. Research}\ }\textbf {\bibinfo {volume} {4}},\ \bibinfo
  {pages} {023089} (\bibinfo {year} {2022})}\BibitemShut {NoStop}%
\bibitem [{\citenamefont {Wang}\ \emph {et~al.}(2021)\citenamefont {Wang},
  \citenamefont {Dutt}, \citenamefont {Yang}, \citenamefont {Wojcik},
  \citenamefont {Vu{\v{c}}kovi{\'{c}}},\ and\ \citenamefont {Fan}}]{Wang_2021}%
  \BibitemOpen
  \bibfield  {author} {\bibinfo {author} {\bibfnamefont {Kai}\ \bibnamefont
  {Wang}}, \bibinfo {author} {\bibfnamefont {Avik}\ \bibnamefont {Dutt}},
  \bibinfo {author} {\bibfnamefont {Ki~Youl}\ \bibnamefont {Yang}}, \bibinfo
  {author} {\bibfnamefont {Casey~C.}\ \bibnamefont {Wojcik}}, \bibinfo {author}
  {\bibfnamefont {Jelena}\ \bibnamefont {Vu{\v{c}}kovi{\'{c}}}}, \ and\
  \bibinfo {author} {\bibfnamefont {Shanhui}\ \bibnamefont {Fan}},\ }\bibfield
  {title} {\enquote {\bibinfo {title} {Generating arbitrary topological
  windings of a non-{H}ermitian band},}\ }\href {\doibase
  10.1126/science.abf6568} {\bibfield  {journal} {\bibinfo  {journal}
  {Science}\ }\textbf {\bibinfo {volume} {371}},\ \bibinfo {pages} {1240--1245}
  (\bibinfo {year} {2021})}\BibitemShut {NoStop}%
\bibitem [{\citenamefont {Hatano}\ and\ \citenamefont
  {Nelson}(1997)}]{Hatano1997}%
  \BibitemOpen
  \bibfield  {author} {\bibinfo {author} {\bibfnamefont {Naomichi}\
  \bibnamefont {Hatano}}\ and\ \bibinfo {author} {\bibfnamefont {David~R.}\
  \bibnamefont {Nelson}},\ }\bibfield  {title} {\enquote {\bibinfo {title}
  {Vortex pinning and non-{H}ermitian quantum mechanics},}\ }\href {\doibase
  10.1103/PhysRevB.56.8651} {\bibfield  {journal} {\bibinfo  {journal} {Phys.
  Rev. B}\ }\textbf {\bibinfo {volume} {56}},\ \bibinfo {pages} {8651--8673}
  (\bibinfo {year} {1997})}\BibitemShut {NoStop}%
\bibitem [{\citenamefont {Lodahl}\ \emph {et~al.}(2017)\citenamefont {Lodahl},
  \citenamefont {Mahmoodian}, \citenamefont {Stobbe}, \citenamefont
  {Rauschenbeutel}, \citenamefont {Schneeweiss}, \citenamefont {Volz},
  \citenamefont {Pichler},\ and\ \citenamefont {Zoller}}]{Lodahl2017}%
  \BibitemOpen
  \bibfield  {author} {\bibinfo {author} {\bibfnamefont {Peter}\ \bibnamefont
  {Lodahl}}, \bibinfo {author} {\bibfnamefont {Sahand}\ \bibnamefont
  {Mahmoodian}}, \bibinfo {author} {\bibfnamefont {S{\o}ren}\ \bibnamefont
  {Stobbe}}, \bibinfo {author} {\bibfnamefont {Arno}\ \bibnamefont
  {Rauschenbeutel}}, \bibinfo {author} {\bibfnamefont {Philipp}\ \bibnamefont
  {Schneeweiss}}, \bibinfo {author} {\bibfnamefont {J\"{u}rgen}\ \bibnamefont
  {Volz}}, \bibinfo {author} {\bibfnamefont {Hannes}\ \bibnamefont {Pichler}},
  \ and\ \bibinfo {author} {\bibfnamefont {Peter}\ \bibnamefont {Zoller}},\
  }\bibfield  {title} {\enquote {\bibinfo {title} {Chiral quantum optics},}\
  }\href {\doibase 10.1038/nature21037} {\bibfield  {journal} {\bibinfo
  {journal} {Nature}\ }\textbf {\bibinfo {volume} {541}},\ \bibinfo {pages}
  {473--480} (\bibinfo {year} {2017})}\BibitemShut {NoStop}%
\bibitem [{\citenamefont {Sheremet}\ \emph {et~al.}(2023)\citenamefont
  {Sheremet}, \citenamefont {Petrov}, \citenamefont {Iorsh}, \citenamefont
  {Poshakinskiy},\ and\ \citenamefont {Poddubny}}]{Sheremet2023}%
  \BibitemOpen
  \bibfield  {author} {\bibinfo {author} {\bibfnamefont {Alexandra~S.}\
  \bibnamefont {Sheremet}}, \bibinfo {author} {\bibfnamefont {Mihail~I.}\
  \bibnamefont {Petrov}}, \bibinfo {author} {\bibfnamefont {Ivan~V.}\
  \bibnamefont {Iorsh}}, \bibinfo {author} {\bibfnamefont {Alexander~V.}\
  \bibnamefont {Poshakinskiy}}, \ and\ \bibinfo {author} {\bibfnamefont
  {Alexander~N.}\ \bibnamefont {Poddubny}},\ }\bibfield  {title} {\enquote
  {\bibinfo {title} {Waveguide quantum electrodynamics: Collective radiance and
  photon-photon correlations},}\ }\href {\doibase 10.1103/RevModPhys.95.015002}
  {\bibfield  {journal} {\bibinfo  {journal} {Rev. Mod. Phys.}\ }\textbf
  {\bibinfo {volume} {95}},\ \bibinfo {pages} {015002} (\bibinfo {year}
  {2023})}\BibitemShut {NoStop}%
\bibitem [{\citenamefont {Prasad}\ \emph {et~al.}(2020)\citenamefont {Prasad},
  \citenamefont {Hinney}, \citenamefont {Mahmoodian}, \citenamefont {Hammerer},
  \citenamefont {Rind}, \citenamefont {Schneeweiss}, \citenamefont
  {S{\o}rensen}, \citenamefont {Volz},\ and\ \citenamefont
  {Rauschenbeutel}}]{Prasad2020}%
  \BibitemOpen
  \bibfield  {author} {\bibinfo {author} {\bibfnamefont {Adarsh~S.}\
  \bibnamefont {Prasad}}, \bibinfo {author} {\bibfnamefont {Jakob}\
  \bibnamefont {Hinney}}, \bibinfo {author} {\bibfnamefont {Sahand}\
  \bibnamefont {Mahmoodian}}, \bibinfo {author} {\bibfnamefont {Klemens}\
  \bibnamefont {Hammerer}}, \bibinfo {author} {\bibfnamefont {Samuel}\
  \bibnamefont {Rind}}, \bibinfo {author} {\bibfnamefont {Philipp}\
  \bibnamefont {Schneeweiss}}, \bibinfo {author} {\bibfnamefont {Anders~S.}\
  \bibnamefont {S{\o}rensen}}, \bibinfo {author} {\bibfnamefont {J\"{u}rgen}\
  \bibnamefont {Volz}}, \ and\ \bibinfo {author} {\bibfnamefont {Arno}\
  \bibnamefont {Rauschenbeutel}},\ }\bibfield  {title} {\enquote {\bibinfo
  {title} {Correlating photons using the collective nonlinear response of atoms
  weakly coupled to an optical mode},}\ }\href {\doibase
  10.1038/s41566-020-0692-z} {\bibfield  {journal} {\bibinfo  {journal} {Nature
  Photonics}\ }\textbf {\bibinfo {volume} {14}},\ \bibinfo {pages} {719}
  (\bibinfo {year} {2020})}\BibitemShut {NoStop}%
\bibitem [{\citenamefont {Liedl}\ \emph {et~al.}(2023)\citenamefont {Liedl},
  \citenamefont {Pucher}, \citenamefont {Tebbenjohanns}, \citenamefont
  {Schneeweiss},\ and\ \citenamefont {Rauschenbeutel}}]{Liedl_2023}%
  \BibitemOpen
  \bibfield  {author} {\bibinfo {author} {\bibfnamefont {Christian}\
  \bibnamefont {Liedl}}, \bibinfo {author} {\bibfnamefont {Sebastian}\
  \bibnamefont {Pucher}}, \bibinfo {author} {\bibfnamefont {Felix}\
  \bibnamefont {Tebbenjohanns}}, \bibinfo {author} {\bibfnamefont {Philipp}\
  \bibnamefont {Schneeweiss}}, \ and\ \bibinfo {author} {\bibfnamefont {Arno}\
  \bibnamefont {Rauschenbeutel}},\ }\bibfield  {title} {\enquote {\bibinfo
  {title} {Collective radiation of a cascaded quantum system: From timed
  {D}icke states to inverted ensembles},}\ }\href {\doibase
  10.1103/physrevlett.130.163602} {\bibfield  {journal} {\bibinfo  {journal}
  {Phys. Rev. Lett.}\ }\textbf {\bibinfo {volume} {130}},\ \bibinfo {pages}
  {163602} (\bibinfo {year} {2023})}\BibitemShut {NoStop}%
\bibitem [{\citenamefont {Du}\ \emph {et~al.}(2023)\citenamefont {Du},
  \citenamefont {Guo}, \citenamefont {Zhang},\ and\ \citenamefont
  {Kockum}}]{du2023giant}%
  \BibitemOpen
  \bibfield  {author} {\bibinfo {author} {\bibfnamefont {Lei}\ \bibnamefont
  {Du}}, \bibinfo {author} {\bibfnamefont {Lingzhen}\ \bibnamefont {Guo}},
  \bibinfo {author} {\bibfnamefont {Yan}\ \bibnamefont {Zhang}}, \ and\
  \bibinfo {author} {\bibfnamefont {Anton~Frisk}\ \bibnamefont {Kockum}},\
  }\href@noop {} {\enquote {\bibinfo {title} {Giant emitters in a structured
  bath with non-{H}ermitian skin effect},}\ } (\bibinfo {year} {2023}),\
  \Eprint {http://arxiv.org/abs/2308.16148} {arXiv:2308.16148 [quant-ph]}
  \BibitemShut {NoStop}%
\bibitem [{\citenamefont {Kornovan}\ \emph {et~al.}(2017)\citenamefont
  {Kornovan}, \citenamefont {Petrov},\ and\ \citenamefont
  {Iorsh}}]{Kornovan2017}%
  \BibitemOpen
  \bibfield  {author} {\bibinfo {author} {\bibfnamefont {D.~F.}\ \bibnamefont
  {Kornovan}}, \bibinfo {author} {\bibfnamefont {M.~I.}\ \bibnamefont
  {Petrov}}, \ and\ \bibinfo {author} {\bibfnamefont {I.~V.}\ \bibnamefont
  {Iorsh}},\ }\bibfield  {title} {\enquote {\bibinfo {title} {Transport and
  collective radiance in a basic quantum chiral optical model},}\ }\href
  {\doibase 10.1103/PhysRevB.96.115162} {\bibfield  {journal} {\bibinfo
  {journal} {Phys. Rev. B}\ }\textbf {\bibinfo {volume} {96}},\ \bibinfo
  {pages} {115162} (\bibinfo {year} {2017})}\BibitemShut {NoStop}%
\bibitem [{\citenamefont {Fedorovich}\ \emph {et~al.}(2022)\citenamefont
  {Fedorovich}, \citenamefont {Kornovan}, \citenamefont {Poddubny},\ and\
  \citenamefont {Petrov}}]{Fedorovich2022}%
  \BibitemOpen
  \bibfield  {author} {\bibinfo {author} {\bibfnamefont {Gleb}\ \bibnamefont
  {Fedorovich}}, \bibinfo {author} {\bibfnamefont {Danil}\ \bibnamefont
  {Kornovan}}, \bibinfo {author} {\bibfnamefont {Alexander}\ \bibnamefont
  {Poddubny}}, \ and\ \bibinfo {author} {\bibfnamefont {Mihail}\ \bibnamefont
  {Petrov}},\ }\bibfield  {title} {\enquote {\bibinfo {title} {Chirality-driven
  delocalization in disordered waveguide-coupled quantum arrays},}\ }\href
  {\doibase 10.1103/PhysRevA.106.043723} {\bibfield  {journal} {\bibinfo
  {journal} {Phys. Rev. A}\ }\textbf {\bibinfo {volume} {106}},\ \bibinfo
  {pages} {043723} (\bibinfo {year} {2022})}\BibitemShut {NoStop}%
\bibitem [{\citenamefont {Yang}\ \emph {et~al.}(2022)\citenamefont {Yang},
  \citenamefont {Wang}, \citenamefont {Wu}, \citenamefont {Xiao}, \citenamefont
  {Yu}, \citenamefont {Yuan},\ and\ \citenamefont {Chen}}]{Yang_2022}%
  \BibitemOpen
  \bibfield  {author} {\bibinfo {author} {\bibfnamefont {Mengjie}\ \bibnamefont
  {Yang}}, \bibinfo {author} {\bibfnamefont {Luojia}\ \bibnamefont {Wang}},
  \bibinfo {author} {\bibfnamefont {Xiaoxiong}\ \bibnamefont {Wu}}, \bibinfo
  {author} {\bibfnamefont {Han}\ \bibnamefont {Xiao}}, \bibinfo {author}
  {\bibfnamefont {Danying}\ \bibnamefont {Yu}}, \bibinfo {author}
  {\bibfnamefont {Luqi}\ \bibnamefont {Yuan}}, \ and\ \bibinfo {author}
  {\bibfnamefont {Xianfeng}\ \bibnamefont {Chen}},\ }\bibfield  {title}
  {\enquote {\bibinfo {title} {Concentrated subradiant modes in a
  one-dimensional atomic array coupled with chiral waveguides},}\ }\href
  {\doibase 10.1103/physreva.106.043717} {\bibfield  {journal} {\bibinfo
  {journal} {Phys. Rev. A}\ }\textbf {\bibinfo {volume} {106}},\ \bibinfo
  {pages} {043717} (\bibinfo {year} {2022})}\BibitemShut {NoStop}%
\bibitem [{\citenamefont {Wang}\ \emph {et~al.}(2022)\citenamefont {Wang},
  \citenamefont {You},\ and\ \citenamefont {Jen}}]{Wang_2022}%
  \BibitemOpen
  \bibfield  {author} {\bibinfo {author} {\bibfnamefont {Yi-Cheng}\
  \bibnamefont {Wang}}, \bibinfo {author} {\bibfnamefont {Jhih-Shih}\
  \bibnamefont {You}}, \ and\ \bibinfo {author} {\bibfnamefont {H.~H.}\
  \bibnamefont {Jen}},\ }\bibfield  {title} {\enquote {\bibinfo {title} {A
  non-{H}ermitian optical atomic mirror},}\ }\href {\doibase
  10.1038/s41467-022-32372-3} {\bibfield  {journal} {\bibinfo  {journal} {Nat
  Commun}\ }\textbf {\bibinfo {volume} {13}},\ \bibinfo {pages} {4598}
  (\bibinfo {year} {2022})}\BibitemShut {NoStop}%
\bibitem [{\citenamefont {Yu}\ and\ \citenamefont {Zeng}(2022)}]{Zeng2022}%
  \BibitemOpen
  \bibfield  {author} {\bibinfo {author} {\bibfnamefont {Tao}\ \bibnamefont
  {Yu}}\ and\ \bibinfo {author} {\bibfnamefont {Bowen}\ \bibnamefont {Zeng}},\
  }\bibfield  {title} {\enquote {\bibinfo {title} {Giant microwave sensitivity
  of a magnetic array by long-range chiral interaction driven skin effect},}\
  }\href {\doibase 10.1103/PhysRevB.105.L180401} {\bibfield  {journal}
  {\bibinfo  {journal} {Phys. Rev. B}\ }\textbf {\bibinfo {volume} {105}},\
  \bibinfo {pages} {L180401} (\bibinfo {year} {2022})}\BibitemShut {NoStop}%
\bibitem [{\citenamefont {Zeng}\ and\ \citenamefont {Yu}(2023)}]{Zeng2023}%
  \BibitemOpen
  \bibfield  {author} {\bibinfo {author} {\bibfnamefont {Bowen}\ \bibnamefont
  {Zeng}}\ and\ \bibinfo {author} {\bibfnamefont {Tao}\ \bibnamefont {Yu}},\
  }\bibfield  {title} {\enquote {\bibinfo {title} {Radiation-free and
  non-hermitian topology inertial defect states of on-chip magnons},}\ }\href
  {\doibase 10.1103/physrevresearch.5.013003} {\bibfield  {journal} {\bibinfo
  {journal} {Phys. Rev. Research}\ }\textbf {\bibinfo {volume} {5}},\ \bibinfo
  {pages} {013003} (\bibinfo {year} {2023})}\BibitemShut {NoStop}%
\bibitem [{\citenamefont {{Voronov}}\ \emph {et~al.}(2007)\citenamefont
  {{Voronov}}, \citenamefont {{Ivchenko}}, \citenamefont {{Erementchouk}},
  \citenamefont {{Deych}},\ and\ \citenamefont {{Lisyansky}}}]{Voronov2007JLu}%
  \BibitemOpen
  \bibfield  {author} {\bibinfo {author} {\bibfnamefont {M.}~\bibnamefont
  {{Voronov}}}, \bibinfo {author} {\bibfnamefont {E.}~\bibnamefont
  {{Ivchenko}}}, \bibinfo {author} {\bibfnamefont {M.}~\bibnamefont
  {{Erementchouk}}}, \bibinfo {author} {\bibfnamefont {L.}~\bibnamefont
  {{Deych}}}, \ and\ \bibinfo {author} {\bibfnamefont {A.}~\bibnamefont
  {{Lisyansky}}},\ }\bibfield  {title} {\enquote {\bibinfo {title}
  {{P}hotoluminescence spectroscopy of one-dimensional resonant photonic
  crystals},}\ }\href {\doibase 10.1016/j.jlumin.2006.08.015} {\bibfield
  {journal} {\bibinfo  {journal} {J. of Luminescence}\ }\textbf {\bibinfo
  {volume} {125}},\ \bibinfo {pages} {112--117} (\bibinfo {year}
  {2007})}\BibitemShut {NoStop}%
\bibitem [{\citenamefont {Tsakmakidis}\ \emph {et~al.}(2017)\citenamefont
  {Tsakmakidis}, \citenamefont {Shen}, \citenamefont {Schulz}, \citenamefont
  {Zheng}, \citenamefont {Upham}, \citenamefont {Deng}, \citenamefont {Altug},
  \citenamefont {Vakakis},\ and\ \citenamefont {Boyd}}]{Tsakmakidis_2017}%
  \BibitemOpen
  \bibfield  {author} {\bibinfo {author} {\bibfnamefont {K.~L.}\ \bibnamefont
  {Tsakmakidis}}, \bibinfo {author} {\bibfnamefont {L.}~\bibnamefont {Shen}},
  \bibinfo {author} {\bibfnamefont {S.~A.}\ \bibnamefont {Schulz}}, \bibinfo
  {author} {\bibfnamefont {X.}~\bibnamefont {Zheng}}, \bibinfo {author}
  {\bibfnamefont {J.}~\bibnamefont {Upham}}, \bibinfo {author} {\bibfnamefont
  {X.}~\bibnamefont {Deng}}, \bibinfo {author} {\bibfnamefont {H.}~\bibnamefont
  {Altug}}, \bibinfo {author} {\bibfnamefont {A.~F.}\ \bibnamefont {Vakakis}},
  \ and\ \bibinfo {author} {\bibfnamefont {R.~W.}\ \bibnamefont {Boyd}},\
  }\bibfield  {title} {\enquote {\bibinfo {title} {Breaking {L}orentz
  reciprocity to overcome the time-bandwidth limit in physics and
  engineering},}\ }\href {\doibase 10.1126/science.aam6662} {\bibfield
  {journal} {\bibinfo  {journal} {Science}\ }\textbf {\bibinfo {volume}
  {356}},\ \bibinfo {pages} {1260--1264} (\bibinfo {year} {2017})}\BibitemShut
  {NoStop}%
\bibitem [{\citenamefont {Mann}\ \emph {et~al.}(2019)\citenamefont {Mann},
  \citenamefont {Sounas},\ and\ \citenamefont {Al\`{u}}}]{Mann2019}%
  \BibitemOpen
  \bibfield  {author} {\bibinfo {author} {\bibfnamefont {Sander~A.}\
  \bibnamefont {Mann}}, \bibinfo {author} {\bibfnamefont {Dimitrios~L.}\
  \bibnamefont {Sounas}}, \ and\ \bibinfo {author} {\bibfnamefont {Andrea}\
  \bibnamefont {Al\`{u}}},\ }\bibfield  {title} {\enquote {\bibinfo {title}
  {Nonreciprocal cavities and the time--bandwidth limit},}\ }\href {\doibase
  10.1364/OPTICA.6.000104} {\bibfield  {journal} {\bibinfo  {journal} {Optica}\
  }\textbf {\bibinfo {volume} {6}},\ \bibinfo {pages} {104--110} (\bibinfo
  {year} {2019})}\BibitemShut {NoStop}%
\bibitem [{\citenamefont {Tsakmakidis}\ \emph {et~al.}(2020)\citenamefont
  {Tsakmakidis}, \citenamefont {You}, \citenamefont {Stefa{\'{n}}ski},\ and\
  \citenamefont {Shen}}]{Tsakmakidis2020}%
  \BibitemOpen
  \bibfield  {author} {\bibinfo {author} {\bibfnamefont {Kosmas~L.}\
  \bibnamefont {Tsakmakidis}}, \bibinfo {author} {\bibfnamefont {Yun}\
  \bibnamefont {You}}, \bibinfo {author} {\bibfnamefont {Tomasz}\ \bibnamefont
  {Stefa{\'{n}}ski}}, \ and\ \bibinfo {author} {\bibfnamefont {Linfang}\
  \bibnamefont {Shen}},\ }\bibfield  {title} {\enquote {\bibinfo {title}
  {Nonreciprocal cavities and the time-bandwidth limit: comment},}\ }\href
  {\doibase 10.1364/optica.384840} {\bibfield  {journal} {\bibinfo  {journal}
  {Optica}\ }\textbf {\bibinfo {volume} {7}},\ \bibinfo {pages} {1097}
  (\bibinfo {year} {2020})}\BibitemShut {NoStop}%
\bibitem [{\citenamefont {Mann}\ \emph {et~al.}(2020)\citenamefont {Mann},
  \citenamefont {Sounas},\ and\ \citenamefont {Al{\`{u}}}}]{Mann_2020}%
  \BibitemOpen
  \bibfield  {author} {\bibinfo {author} {\bibfnamefont {Sander~A.}\
  \bibnamefont {Mann}}, \bibinfo {author} {\bibfnamefont {Dimitrios~L.}\
  \bibnamefont {Sounas}}, \ and\ \bibinfo {author} {\bibfnamefont {Andrea}\
  \bibnamefont {Al{\`{u}}}},\ }\bibfield  {title} {\enquote {\bibinfo {title}
  {Nonreciprocal cavities and the time-bandwidth limit: reply},}\ }\href
  {\doibase 10.1364/optica.401383} {\bibfield  {journal} {\bibinfo  {journal}
  {Optica}\ }\textbf {\bibinfo {volume} {7}},\ \bibinfo {pages} {1102}
  (\bibinfo {year} {2020})}\BibitemShut {NoStop}%
\bibitem [{\citenamefont {Buddhiraju}\ \emph {et~al.}(2020)\citenamefont
  {Buddhiraju}, \citenamefont {Shi}, \citenamefont {Song}, \citenamefont
  {Wojcik}, \citenamefont {Minkov}, \citenamefont {Williamson}, \citenamefont
  {Dutt},\ and\ \citenamefont {Fan}}]{Buddhiraju_2020}%
  \BibitemOpen
  \bibfield  {author} {\bibinfo {author} {\bibfnamefont {Siddharth}\
  \bibnamefont {Buddhiraju}}, \bibinfo {author} {\bibfnamefont
  {Yu}~\bibnamefont {Shi}}, \bibinfo {author} {\bibfnamefont {Alex}\
  \bibnamefont {Song}}, \bibinfo {author} {\bibfnamefont {Casey}\ \bibnamefont
  {Wojcik}}, \bibinfo {author} {\bibfnamefont {Momchil}\ \bibnamefont
  {Minkov}}, \bibinfo {author} {\bibfnamefont {Ian A.~D.}\ \bibnamefont
  {Williamson}}, \bibinfo {author} {\bibfnamefont {Avik}\ \bibnamefont {Dutt}},
  \ and\ \bibinfo {author} {\bibfnamefont {Shanhui}\ \bibnamefont {Fan}},\
  }\bibfield  {title} {\enquote {\bibinfo {title} {Absence of unidirectionally
  propagating surface plasmon-polaritons at nonreciprocal metal-dielectric
  interfaces},}\ }\href {\doibase 10.1038/s41467-020-14504-9} {\bibfield
  {journal} {\bibinfo  {journal} {Nat Commun}\ }\textbf {\bibinfo {volume}
  {11}},\ \bibinfo {pages} {674} (\bibinfo {year} {2020})}\BibitemShut
  {NoStop}%
\bibitem [{\citenamefont {Wang}\ \emph {et~al.}(2023)\citenamefont {Wang},
  \citenamefont {Jen},\ and\ \citenamefont {You}}]{Wang_2023}%
  \BibitemOpen
  \bibfield  {author} {\bibinfo {author} {\bibfnamefont {Yi-Cheng}\
  \bibnamefont {Wang}}, \bibinfo {author} {\bibfnamefont {H.~H.}\ \bibnamefont
  {Jen}}, \ and\ \bibinfo {author} {\bibfnamefont {Jhih-Shih}\ \bibnamefont
  {You}},\ }\bibfield  {title} {\enquote {\bibinfo {title} {Scaling laws for
  non-{H}ermitian skin effect with long-range couplings},}\ }\href {\doibase
  10.1103/physrevb.108.085418} {\bibfield  {journal} {\bibinfo  {journal}
  {Phys. Rev. B}\ }\textbf {\bibinfo {volume} {108}},\ \bibinfo {pages}
  {085418} (\bibinfo {year} {2023})}\BibitemShut {NoStop}%
\bibitem [{\citenamefont {Berestetskii}\ \emph {et~al.}(1982)\citenamefont
  {Berestetskii}, \citenamefont {Lifshitz},\ and\ \citenamefont
  {Pitaevskii}}]{berestetskii1982quantum}%
  \BibitemOpen
  \bibfield  {author} {\bibinfo {author} {\bibfnamefont {V.B.}\ \bibnamefont
  {Berestetskii}}, \bibinfo {author} {\bibfnamefont {E.M.}\ \bibnamefont
  {Lifshitz}}, \ and\ \bibinfo {author} {\bibfnamefont {L.P.}\ \bibnamefont
  {Pitaevskii}},\ }\href@noop {} {\emph {\bibinfo {title} {Quantum
  Electrodynamics: Volume 4}}},\ Vol.~\bibinfo {volume} {4}\ (\bibinfo
  {publisher} {Butterworth-Heinemann},\ \bibinfo {year} {1982})\BibitemShut
  {NoStop}%
\bibitem [{\citenamefont {Cardenas-Lopez}\ \emph {et~al.}(2023)\citenamefont
  {Cardenas-Lopez}, \citenamefont {Masson}, \citenamefont {Zager},\ and\
  \citenamefont {Asenjo-Garcia}}]{Cardenas_Lopez_2023}%
  \BibitemOpen
  \bibfield  {author} {\bibinfo {author} {\bibfnamefont {Silvia}\ \bibnamefont
  {Cardenas-Lopez}}, \bibinfo {author} {\bibfnamefont {Stuart~J.}\ \bibnamefont
  {Masson}}, \bibinfo {author} {\bibfnamefont {Zoe}\ \bibnamefont {Zager}}, \
  and\ \bibinfo {author} {\bibfnamefont {Ana}\ \bibnamefont {Asenjo-Garcia}},\
  }\bibfield  {title} {\enquote {\bibinfo {title} {Many-body superradiance and
  dynamical mirror symmetry breaking in waveguide {QED}},}\ }\href {\doibase
  10.1103/physrevlett.131.033605} {\bibfield  {journal} {\bibinfo  {journal}
  {Phys. Rev. Lett.}\ }\textbf {\bibinfo {volume} {131}},\ \bibinfo {pages}
  {033605} (\bibinfo {year} {2023})}\BibitemShut {NoStop}%
\bibitem [{\citenamefont {Kivshar}(2022)}]{Kivshar2022}%
  \BibitemOpen
  \bibfield  {author} {\bibinfo {author} {\bibfnamefont {Yuri}\ \bibnamefont
  {Kivshar}},\ }\bibfield  {title} {\enquote {\bibinfo {title} {Mie scattering
  yields chiral nonlinearity},}\ }\href {\doibase 10.1038/s41566-022-00953-9}
  {\bibfield  {journal} {\bibinfo  {journal} {Nature Photonics}\ }\textbf
  {\bibinfo {volume} {16}},\ \bibinfo {pages} {89--90} (\bibinfo {year}
  {2022})}\BibitemShut {NoStop}%
\bibitem [{\citenamefont {Koshelev}\ \emph {et~al.}(2023)\citenamefont
  {Koshelev}, \citenamefont {Tang}, \citenamefont {Hu}, \citenamefont
  {Kravchenko}, \citenamefont {Li},\ and\ \citenamefont
  {Kivshar}}]{Koshelev2023}%
  \BibitemOpen
  \bibfield  {author} {\bibinfo {author} {\bibfnamefont {Kirill}\ \bibnamefont
  {Koshelev}}, \bibinfo {author} {\bibfnamefont {Yutao}\ \bibnamefont {Tang}},
  \bibinfo {author} {\bibfnamefont {Zixian}\ \bibnamefont {Hu}}, \bibinfo
  {author} {\bibfnamefont {Ivan~I.}\ \bibnamefont {Kravchenko}}, \bibinfo
  {author} {\bibfnamefont {Guixin}\ \bibnamefont {Li}}, \ and\ \bibinfo
  {author} {\bibfnamefont {Yuri}\ \bibnamefont {Kivshar}},\ }\bibfield  {title}
  {\enquote {\bibinfo {title} {Resonant chiral effects in nonlinear dielectric
  metasurfaces},}\ }\href {\doibase 10.1021/acsphotonics.2c01926} {\bibfield
  {journal} {\bibinfo  {journal} {{ACS} Photonics}\ }\textbf {\bibinfo {volume}
  {10}},\ \bibinfo {pages} {298--306} (\bibinfo {year} {2023})}\BibitemShut
  {NoStop}%
\end{thebibliography}
%

\end{document}